\documentclass[useAMS,usenatbib]{mn2e}
  \def\ark{Ark 120}
  \def\pks{PKS~0558--504}

  \def\xmm{{\it XMM-Newton}}

  \def\swift{{\it SWIFT}}

  \def\ltsima{$\; \buildrel < \over \sim \;$}
  \def\simlt{\lower.5ex\hbox{\ltsima}} 
  \def\gtsima{$\; \buildrel > \over \sim \;$}
  \def\simgt{\lower.5ex\hbox{\gtsima}} 
  \def\plm{$\pm$}
\usepackage{graphics,graphicx}
\usepackage{epsfig}
\usepackage{dcolumn}
\usepackage{rotating}

\title[Swift monitoring of Ark 120]{Long-term monitoring of Ark 120 with Swift}
\author[M. Gliozzi et al.]{M. Gliozzi,$^{1}$\thanks{E-mail: mgliozzi@gmu.edu},
I.E. Papadakis,$^{2,3}$ D. Grupe,$^{4}$ W.P. Brinkmann,$^{5}$ and C. R\"ath$^{6}$\\
$^{1}$ Physics and Astronomy Department,
George Mason University, 4400 University Drive, Fairfax, VA 22030\\
$^{2}$ Physics Department, University of Crete, 710 03 Heraklion,
Crete, Greece\\
$^{3}$ Foundation for Research and Technology - Hellas,
IESL, Voutes, 71110 Heraklion, Crete, Greece\\
$^{4}$ Department of Earth and Space Sciences, Morehead State University, Morehead, Kentucky, USA\\
$^{5}$ Max-Planck-Institut f\"ur 
extraterrestrische Physik, Postfach 1312, D-85741 Garching, Germany\\
$^{5}$ German Aerospace Center, Department of Complex Plasmas, K\"oln, Germany}
\begin{document}
%
%
%
\maketitle
\begin{abstract}
We report the results of a six month \swift\ monitoring campaign of Ark 120, a 
prototypical ``bare" Seyfert 1 galaxy. The lack of intrinsic absorption combined
with the nearly contemporaneous coverage of the UV, and X-ray bands 
makes it possible to investigate the link between the accretion disk and the 
putative Comptonization corona. Our observations confirm the presence of 
substantial temporal variability, with the X-ray characterized by large-amplitude
flux changes on timescales of few days, while the variations in the UV bands are
smoother and occur on timescales of several weeks. The source also shows spectral variability with the X-ray spectrum steepening when the source is brighter. We do not detect any correlation between the UV flux and the X-ray spectral slope. 
A cross correlation analysis suggests positive delays between X-rays and the UV emission, favoring a scenario of disk reprocessing. Although the strength of the correlation is moderate with a delay that is not well constrained ($7.5\pm7$ days), it is nevertheless indicative of a very large disk reprocessing region, with a separation between the X-ray and the UV emitting regions which could be as large as 1000 $r_G$. The Ark 120 correlation results are in agreement with those obtained in similar multi-wavelength monitoring studies of AGN. When combined together, the observations so far can be well described by a linear relation between the X-ray/UV delays and the mass of the central black hole. Within the context of the simplest scenario where these delays correspond to light-travel times, the implied distance between the X-ray source and the optical/UV disk reprocessing region in these AGN should be of the order of many hundreds of gravitational radii.
\end{abstract}
\section{Introduction}
Numerous observational studies across the electromagnetic spectrum have shown 
that active galactic nuclei (AGN) are powerful emitters
of variable radiation from the radio band to $\gamma$-ray energies and that 
a sizable amount of this radiation is emitted in the optical, UV, X-ray energy bands.
It is generally thought that AGN are powered by accretion onto a central 
supermassive black hole and that the optical-UV emission is the thermal radiation 
produced directly from the accretion flow, whereas the X-rays are produced through
the Comptonization process in a putative corona. While this general picture
is widely accepted, the details of the interaction between disk
and corona, the geometry and physical state of the latter, as well as the origin of the variability are still poorly understood.

One of the most promising approaches to shed some light on the inner region of black
holes systems
is to track the variable behavior of the different components of the central engine
using multi-wavelength monitoring observations. Over the years, coordinated optical 
and X-ray monitoring observations of AGN have revealed that,
in addition to the ubiquitous X-ray variability, long-term optical variations are
consistently detected. The latter are generally interpreted
as intrinsic variations of the disk emission caused by disturbances
propagating across the disk or, alternatively, as a result of the disk
reprocessing of the X-ray emission produced by the corona. In this context,
various monitoring studies have yielded contrasting results, with some sources showing 
highly correlated optical/X-ray behavior with one band leading the other, and some 
others showing no correlation at all 
\citep[e.g.,][]{maoz00,utt03,arev09,bree09,arev09,bree10}. Some of the inferred discrepancies may be attributed
to the intrinsic difficulties of coordinating satellite-based X-ray monitoring with
ground-based optical observations, which may also be hampered by weather conditions.

An important step forward in this field has been the advent of \swift, which, thanks to its flexibility and simultaneous 
coverage of several bands in the optical/UV range and in the X-rays, eliminates
all possible ambiguities associated with weather conditions and lack of coordination in the different energy bands. In principle, true contemporaneous observations in several energy bands should make it possible to test wether the diverse temporal behavior observed in different objects is related to some fundamental properties of the central engine. More specifically, different combinations of  
black hole mass $M_{\rm BH}$ and accretion rate $\dot m$ may be responsible for the 
diverse behavior observed in various BH systems. For example, Uttley
et al. (2003) hypothesized that objects with large $M_{\rm BH}$ and low $\dot m$ 
should yield a relatively tight X-ray/optical correlation, since they
have cooler disks and hence the optical region closer (in units of gravitational 
radii) to the X-ray emitting central corona. Conversely, for AGN with small 
$M_{\rm BH}$ and high $\dot m$, the region producing the optical is relatively
distant from the X-ray region (the corona) and hence an uncorrelated optical/X-ray
behavior may be expected.

In our previous work, we investigated with \swift\ the long-term 
behavior of \pks, an X-ray bright radio-loud Narrow-line Seyfert 1 (NLS1)
with $M_{\rm BH}\sim 3\times10^8~{M_\odot}$ likely accreting at super-Eddington rate.
The main findings of our 1.5 year monitoring campaign can be summarized as follows.
\pks\ is highly variable at all wavelengths probed by \swift,
with variability levels decreasing from the X-rays to
the optical bands.  The large-amplitude variations
measured by the UVW2 filter are strongly correlated with all
the other optical and UV bands and weakly (but significantly) correlated with the
X-ray variations. These results, combined with suggestive evidence that in \pks\
perturbations propagate from the outer to the inner parts of the accretion flow
and to the corona, confirm the existence of physical link between 
disk and corona and disfavor the reprocessing scenario (for more details see \cite{glioz13}).

Here, to further investigate the link between accretion disk and corona and its
putative dependence on $M_{\rm BH}$ and $\dot m$,
we study the correlated variability properties of the prototypical bare Seyfert 1
galaxy \ark, which has a relatively large black hole mass, 
$M_{\rm BH}\sim 1.5\times10^8~{M_\odot}$ \citep{peter04} and 
low accretion rate $L_{\rm bol}/L_{\rm Edd} \sim 0.05$ \citep{vasu07}. Past studies have shown that \ark\ stands out among bright Seyfert galaxies for the lack of any evidence
for reddening in IR and UV observations \citep{war87,cren99} and warm absorbers 
\citep{vaugh04} and is highly variable source over the entire spectrum 
\citep[e.g.,][]{kas00,cari03,doro08}, making it
an ideal target for a \swift\ monitoring campaign.

This paper is structured as follows. In Section 2, we describe the observations and data reduction. In Section 3 
we study the temporal and spectral variability properties with model-independent tools, in Section 4 we describe the results of a standard spectral analysis, whereas in Section 5 we carry out an inter-band correlation analysis. 
In Section 6 we summarize our main findings and discuss their implications.

Hereafter, we adopt a cosmology with $H_0=71{\rm~km~s^{-1}~Mpc^{-1}}$,
$\Omega_\Lambda=0.73$ and $\Omega_{\rm M}=0.27$ \citep{ben03}. 
With the assumed cosmological parameters, the luminosity
distance of \ark\  is 142 Mpc.

\section{Observations and Data Reduction}
\ark\ was observed by \swift\ \citep{ger04} between 4 September 
2014 and 9 March 2015 with a cadence of one pointing every two days. 
This observing strategy was devised to preserve the UVOT filter wheel by using
the filters of the day U and UVM2, which regularly alternate every two days. 
The filters choice is based on the fact that U is the ``reddest" among 
standard filters and 
UVM2 the ``bluest" clean filter (UVW2 is slightly bluer than UVM2
with a central wavelength of 1950 \AA\ vs. 2200\AA, but is affected by a red leak).
The details of the \swift\ monitoring campaign are summarized in Table 1  (dates and exposures) and Table 2 (X-ray count rates and hardness ratios, as well as U and UVM2
fluxes), where only the first 
four entries are shown for illustrative purposes. The complete tables are available 
in electronic format. The \swift\ XRT \citep{burro05} 
observations were performed in windowed timing mode  
to avoid possible pile-up effects (\citealt{hill04}). 
For both spectra and light curves, source photons were extracted in a circular region 
with a radius of 20 pixels (corresponding to $\sim$47\arcsec) centered on the source, 
and the background was selected using a circle of the same size but shifted along 
the window, away from the source position. Only single to 
quadruple events in the energy range of 0.3--10 keV were selected for further
analysis. Source and background spectra were extracted from the event file 
by using XSELECT version 2.3. All the light curves analyzed in Section 3 are background-subtracted.
The auxiliary response files were created by the Swift tool {\tt xrtmkarf} and used in combination with
the response matrix \texttt{swxwt0to2s6psf1\_20131212v001.rmf}. 
We also obtained photometry with the UV/Optical 
Telescope (UVOT; \citealt{pool08}, \citealt{breev10}) in the U and UVM2. Source photons were 
extracted from a circular region with r = 5\arcsec, and the background from an 
source-free circular region with a 
radius of 20\arcsec. The UVOT tool {\tt uvotsource} was used to determine the 
fluxes.  The fluxes were corrected for Galactic reddening
($E_{\rm B-V}$=0.113 obtained from NED) with the standard
reddening correction curves by \citet{card89}. The U and UVM2 images of \ark\ show point-like sources, suggesting that the galaxy contribution is negligible compared to the AGN emission. Indeed, this conclusion is confirmed by \citet{koss11}, who studied the properties of the host galaxies of the AGN observed with \swift\ BAT and concluded that the galaxy emission of \ark\ is strongly contaminated by the AGN in all the filters.
\setcounter{table}{0}
\begin{table*} 
\centering
\begin{minipage}{180mm}
\caption{Observation log of Ark 120}
\begin{tabular}{lccccccc} 
\hline        
\hline
\noalign{\smallskip} 
Segment & Start time (UT) & End Time (UT) & MJD & \multicolumn{3}{c}{Observing time given in s}\\
 & & & & $\rm T_{\rm XRT}$& $\rm T_{\rm U}$  & $\rm T_{\rm UVM2}$\\
\noalign{\smallskip}
\hline
\noalign{\smallskip}
01 & 2014-09-04 01:20:00 & 2014-09-04 01:36:59 & 56904.56 & 992 & 998 & \dots \\
02 & 2014-09-06 04:34:56 & 2014-09-06 04:51:58 & 56906.69 & 1009 & \dots & 1000  \\
03 & 2014-09-08 06:03:36 & 2014-09-08 06:20:58 & 56908.75 & 1026 & 1019 &  \dots  \\
04 & 2014-09-10 10:50:37 & 2014-09-10 11:07:58 & 56910.95 & 1038  & \dots & 1018  \\
\noalign{\smallskip}
\hline        
\hline
\end{tabular}
\end{minipage}
\label{tab1}
\end{table*}

\setcounter{table}{1}
\begin{table} 
\begin{minipage}{180mm}
\caption{\swift~XRT count rates and HR and UVOT fluxes (mJy) of Ark 120}
\begin{tabular}{lcccc}
\hline        
\hline
\noalign{\smallskip}
Segment & XRT rate & XRT HR & U & UVM2\\
\noalign{\smallskip}
\hline
\noalign{\smallskip}
01 & 1.19\plm0.03 & 0.20\plm0.02  & 6.93\plm0.15 & \dots      \\
02 & 1.25\plm0.04 & 0.11\plm0.02  & \dots              & 3.57\plm0.08\\
03 & 1.13\plm0.03 & 0.15\plm0.02  & 6.96\plm0.15 &  \dots      \\
04 & 1.31\plm0.04 & 0.18\plm0.02  & \dots              & 3.48\plm0.08\\
\noalign{\smallskip}
\hline        
\hline
\end{tabular}
\end{minipage}
\label{tab2}
\vskip 0.1cm
$HR=(h-s)/(h+s)$, with $s=$0.3-1 keV and $h=$1-10 keV.\\ 
The errors given in this table are statistical errors.
\end{table}

\begin{figure}
\begin{center}
\includegraphics[bb=80 25 470 520,clip=,angle=0,width=8.5cm]{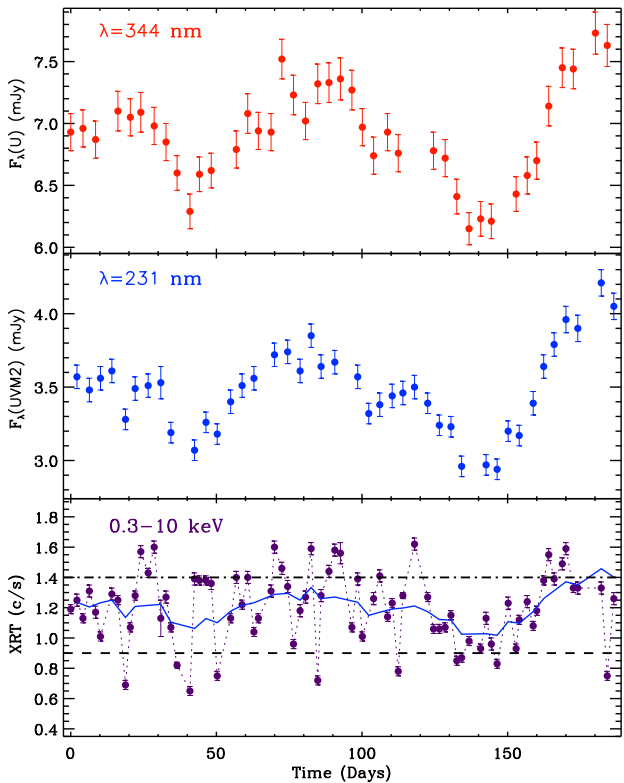}
\caption{\swift\ UVOT U (top panel), UVM2 (middle panel) and 0.3--10 keV XRT (bottom panel) light curves of \ark\ from 4 September 2014 to 9 March 2015. The UV fluxes are corrected for Galactic absorption and expressed in units of mJy. On top of the X-ray time series we superimposed the UVM2 light curve normalized to have the same mean as the X-ray
light curve.
The dot-dashed and dashed lines on the X-ray light curve represent the threshold values used for the flux-selected spectral analysis.} 
\label{figure:fig1}
\end{center}
\end{figure}
\section{Variability analysis of \ark}
Figure~\ref{figure:fig1} shows the UVOT U ($\lambda=344$ nm; top panel), UVM2 ($\lambda=231$ nm; middle  panel) light curves, as well as the 0.3--10 keV XRT light curve (bottom panel) of \ark. On top of the X-ray time series we superimposed the normalized UVM2 light curve (blue continuous line) to highlight the similarities and differences between the temporal behavior in the UV and X-rays. While the overall trend seems to be broadly consistent, the X-ray light
 curve exhibits short-term variability, which is not present in the smoother UV light curves. The dot-dashed and dashed lines on the X-ray light curve represent the threshold values used for the flux-selected spectral analysis (see Section 4); data points above the dot-dashed line have been selected for the ``high-flux" spectrum, whereas those below the dashed line define the ``low-flux" spectrum.

As suggested by the visual inspection of this figure, \ark\ is highly variable at all wavelengths probed by \swift. According to a $\chi^2$ test, the light curves show significant variability with $\chi^2/dof$ values of 270.5/40, 562.6/41, and 4283.8/80 for the U, UVM2 UVOT filters, and XRT, respectively. An analysis of the fractional variability $F_{\rm
var}=\sqrt{\sigma^2-\Delta^2}/\langle r\rangle$ (where $\sigma^2$ is the variance, 
$\Delta^2$ the mean square value of the uncertainty associated with each individual count 
rate, and $\langle r\rangle$ the unweighted mean count rate) confirms these results, suggesting the presence of a positive trend between variability and energy band: 
$F_{\rm var,U}=(5.2\pm0.4)$\%, $F_{\rm var,UVM2}=(7.9\pm0.4)$\%, $F_{\rm var,XRT}=(19.6\pm0.3)$\%.

\begin{figure}
\begin{center}
\includegraphics[bb=80 25 470 510,clip=,angle=0,width=8.5cm]{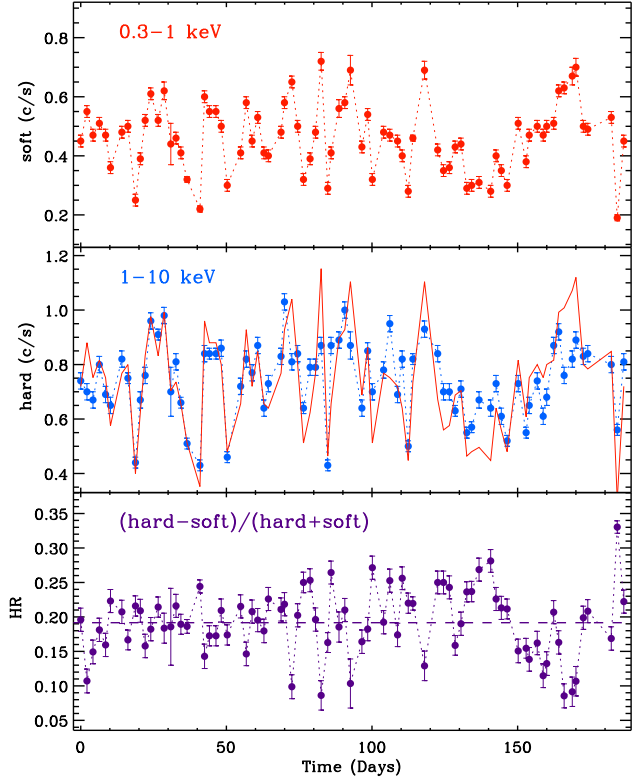}
\caption{Soft, hard, and hardness ratio, HR=(h-s)/(h+s) light curves \ark. The solid line superimposed on the middle panel represents the normalized soft X-ray light curve.
The dashed line in the bottom panel describes the average value of the hardness ratio.} 
\label{figure:fig2}
\end{center}
\end{figure}
Figure~\ref{figure:fig2} shows the soft (0.3--1 keV), hard (1--10 keV), and hardness ratio $HR=(h-s)/(h+s)$ light curves of \ark. To guide the eye, we have superimposed the normalized soft X-ray light curve (red continuous line in the middle panel) on top of the hard
X-ray light curve. The similar trend between the two X-ray bands indicates that 
soft and hard time series vary significantly and nearly in concert. 
According to a  $\chi^2$ test and fractional variability analysis, the count rate variation are highly significant with
$\chi^2/dof=2693.5/80$, $F_{\rm var,soft}=(24.9\pm0.5$)\%, and $4283.8/80$, $F_{\rm var,hard}=(17.9\pm0.4)$\% for the soft and hard band respectively. 
Based on the same tests, the variability
of the $HR$ light curve appears to be statistically significant with $\chi^2/dof=801.3/80$
and $F_{\rm var,HR}=(23\pm1)$\%, suggesting the presence of X-ray spectral variability.

\begin{figure}
\begin{center}
\includegraphics[bb=45 32 350 300,clip=,angle=0,width=8.5cm]{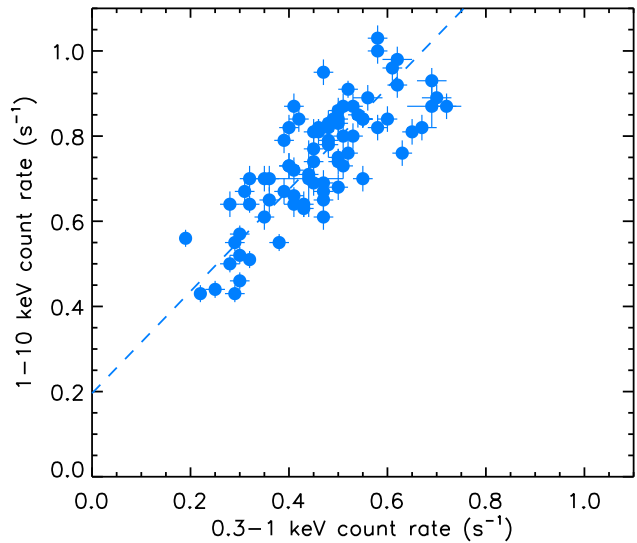}
\caption{Hard vs. soft X-ray count rate plot of \ark\ obtained in the \swift\ XRT campaign. The dashed line represents the best-fit linear model $Count~Rate_{\rm  1-10 keV}= 0.20\pm 0.02 + (1.10\pm 0.04) Count~Rate_{\rm  0.3-1 keV}$.} 
\label{figure:fig3}
\end{center}
\end{figure}
A simple model-independent method to test for spectral variability makes use of the 
plot of the X-ray hard count rate versus the soft count rate \citep[e.g.,][]{chura01}, which is shown in Fig.~\ref{figure:fig3}. As expected from the visual inspection of the X-ray light curves, the plot of hard vs. soft count rates shows a strong linear correlation:  -- the Spearman's rank correlation coefficient is $\rho=0.79$ and associated probability of random correlation $P_\rho=1.4\times10^{-18}$-- and is well described by the equation $Count~Rate_{\rm  1-10 keV}= 0.20\pm 0.02 + (1.10\pm 0.04) Count~Rate_{\rm  0.3-1 keV}$, obtained with the routine {\tt fitexy} \citep{press97}, which accounts for the errors on both  
$y$ and $x$ axes, and will be adopted in the rest
of the paper for any linear correlation analysis. Interestingly, while
the slope is roughly consistent with the unity, the positive intercept, inconsistent with zero, suggests the existence of a non-variable hard component. 

\begin{figure}
\begin{center}
\includegraphics[bb=45 32 350 300,clip=,angle=0,width=8.5cm]{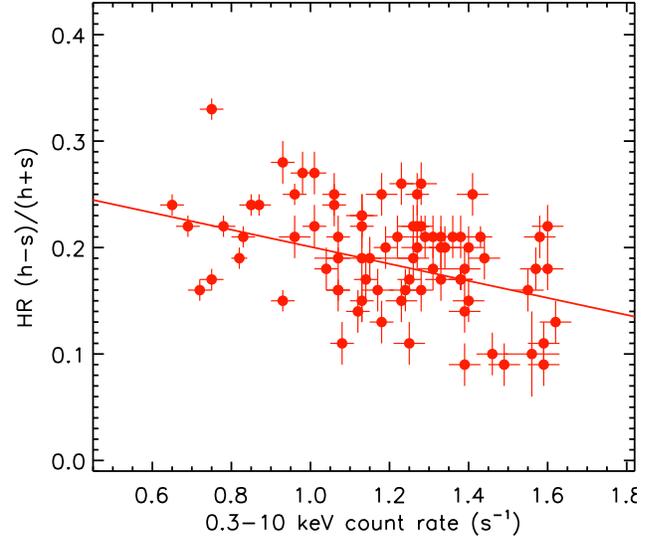}
\caption{Hardness ratio HR=(h-s)/(h+s) vs. 0.3--10 keV X-ray count rate plot of \ark\ obtained in the \swift\ XRT campaign. The continuous line represents the best-fit linear model $HR =  0.287\pm0.008-(0.080\pm 0.007) Count~Rate_{\rm  0.3-10 keV}$.} 
\label{figure:fig4}
\end{center}
\end{figure}
An additional direct way to study the X-ray spectral variability of AGN is to plot the hardness ratio versus the total flux. The result of this analysis for \ark\ is shown in Fig.~\ref{figure:fig4}, which, despite the substantial scatter, reveals the existence
of a shallow but robust anti-correlation ($\rho=-0.38$, $P_\rho=5\times10^{-4}$) described by $HR =  0.287\pm0.008-(0.080\pm 0.007) Count~Rate_{\rm  0.3-10 keV}$, which indicates
that the spectrum softens as the source brightens.

\begin{figure}
\begin{center}
\includegraphics[bb=10 60 590 430,clip=,angle=0,width=8.5cm]{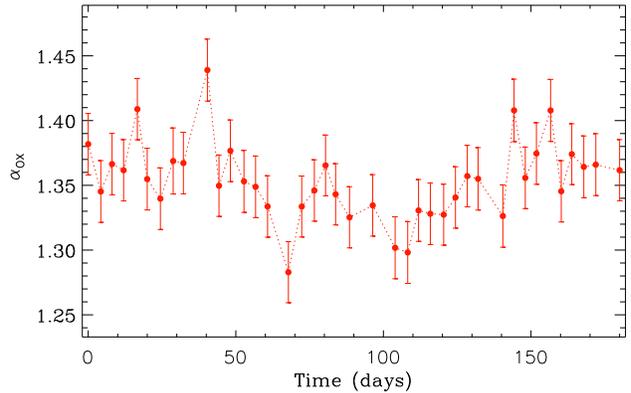}
\caption{Light curve of the broadband spectral index $\alpha_{OX}$.} 
\label{figure:fig5}
\end{center}
\end{figure}
Finally, model-independent information about the broadband spectral variability can be
inferred by studying the temporal evolution of the broadband spectral index 
$\alpha_{\rm OX}=
\log(l_{\rm 2500\AA}/l_{\rm 2keV})/\log(\nu_{\rm 2500\AA}/\nu_{\rm 2keV})$
\citep{tanan79}. We derived $\alpha_{\rm OX}$ from the simultaneous X-ray and 
UVM2 fluxes, and plotted the light curve in Figure~\ref{figure:fig5},
which suggests the presence of a weak variability of the spectral energy distribution (SED): $\chi^2/dof=62.6/39$, $F_{\rm var,\alpha_{\rm OX}}=(1.4\pm0.4)$\%.

In summary, the 6-month \swift\ monitoring campaign of \ark\ confirms the presence of significant large-amplitude variability (for such a large black hole mass) in all bands probed by the UVOT and XRT, with the
X-ray band being by far the most variable component, and indicate that the temporal variability
of \ark\ is associated with spectral changes of the X-rays and, to a lesser extent, of the broadband SED.

\section{X-ray Spectral analysis}
The X-ray spectral analysis  was performed using the {\tt XSPEC v.12.9.0}
software package \citep{arn96}.  For $\sim 1$ ks exposure observations, spectra were 
re-binned within grppha 3.0.0 to 
have at least 1 photon per bin and fitted with the C-statistic, whereas
combined flux-selected spectra were re-binned at 20 counts per channel for the $\chi^2$ statistic to be valid. The errors on spectral parameters represent the 68\% confidence level (1-$\sigma$) for one interesting parameter ($\Delta \chi^2=1$). We verified that the two UV data points are well above the extrapolation of the X-ray best-fitting model and are most likely associated with the accretion disk emission. We did not include the UV data in the spectral fitting analysis, because two non-simultaneous data points in the U and UVM2 filters are not sufficient to characterize the properties of the accretion disk.

We carried out a systematic spectral analysis of every observation of the \ark\ campaign, even though the short exposures of individual \swift\ XRT pointings yield X-ray spectra with limited statistics.  We adopted a baseline model that comprises two Comptonization components (representing the primary emission produced by the corona and the soft excess) and a Gaussian line to account for the iron K$\alpha$ line emission. All additive spectral components are absorbed by a column density fixed at the Galactic value $N_{\rm H}=1.01\times 10^{21}~{\rm cm^{-2}}$, parameterized by the {\tt wabs} model in Xspec. This model choice was guided by the spectral results from past studies with higher signal-to-noise data, 
and more specifically by a recent study based on long exposures from \xmm\ and {\it NuSTAR}, that confirmed the presence of a soft-excess, which appears to be consistent with an additional cooler Comptonization component \citep[]{matt14}. At the beginning all parameters are left free to vary. However, given the limited statistics of the spectra,
some parameters are poorly constrained and yield unreasonable values when computing their statistical uncertainty; in those cases, the parameters are fixed at their best fit value. The parameters left free during the error calculation are the spectral index and the normalization for the individual observations. For the combined spectra of high- and low-flux cases, more parameters are reasonably defined and therefore can be left free to vary during the error calculations.

Both the soft excess and the coronal emission have been parameterized by the Bulk Motion Comptonization (BMC) model in Xspec \citep{tita97}, which is a simple but comprehensive Comptonization model that can fit both
thermal and bulk Comptonization processes, and is described by four parameters: $kT$
(the temperature of the thermal seed photons), $\alpha$ (the energy spectral index 
related to the photon index by the relationship $\Gamma=1+\alpha$), $\log(A)$ (a parameter 
describing to the Comptonization fraction $f=A/(1+A)$), and the normalization.
We used the BMC model instead of the phenomenological
power law model because the BMC parameters are computed in a self-consistent 
way, and the power law produced by BMC does not extend to arbitrarily low energies. 

\begin{figure}
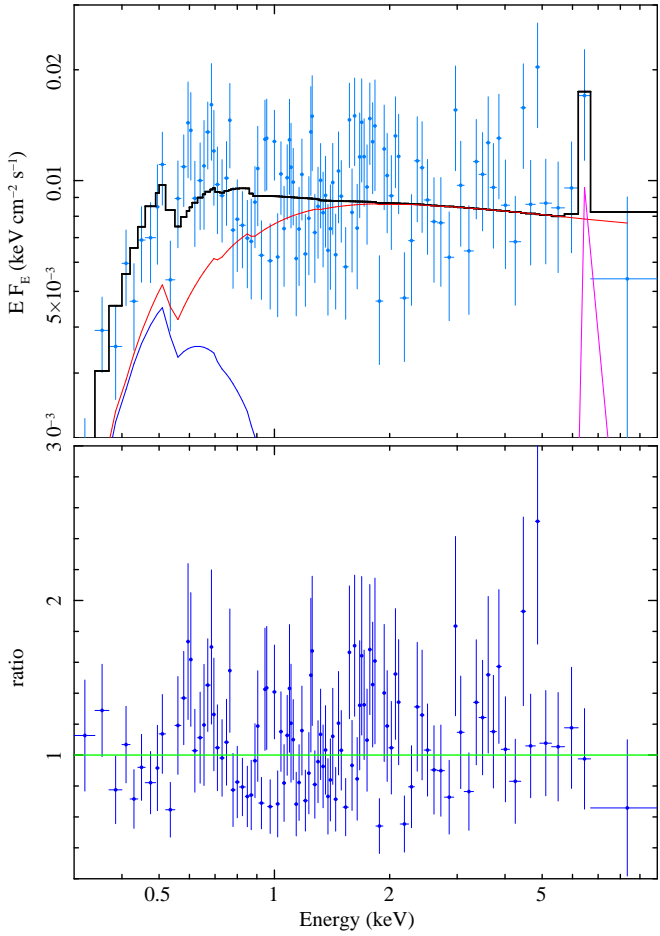

\includegraphics[bb=30 2 530 704,clip=,angle=-90,width=9.cm]{fig6.eps}
\includegraphics[bb=75 2 600 704,clip=,angle=-90,width=9.cm]{fig6b.eps}
\caption{Top panel: Unfolded XRT spectrum of \ark, obtained using the {\tt eeufspec} command in Xspec.  The model includes two BMC components plus one Gaussian line modified by photoelectric absorption. Bottom panel: data to model ratio.}
\label{figure:fig6}
\end{figure}

For illustrative purposes of individual XRT spectra yielded during the \ark\ campaign, the unfolded spectrum ({\tt eeufspec} in Xspec) and the data-to-model ratio from obsid 34 with net exposure of 1039 s and count rate of 
$\sim 1.5$ c/s
are shown in Fig.~\ref{figure:fig6}. This represents one of the best-case scenarios, since it refers to an observation with relatively long exposure and high count rate. Larger uncertainties are associated to observations with shorter exposures or lower count rates. 

\begin{figure}
\begin{center}
\includegraphics[bb=50 60 555 305,clip=,angle=0,width=8.5cm]{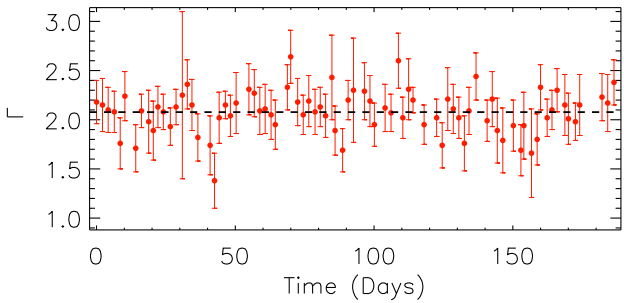}
\caption{Light curve of the X-ray primary emission photon index $\Gamma$ during the \ark\ campaign.} 
\label{figure:fig7}
\end{center}
\end{figure}

All individual observations are reasonably well fitted with this baseline model ($\chi^2_{\rm red}$ ranges from 0.6 to 1.14), although only a few spectra statistically require more than one BMC component, suggesting that the model over-parametrizes the low signal-to-noise spectra. Not surprisingly, the model parameters are poorly constrained. Nevertheless, since the model-independent analysis suggests the presence of spectral variability throughout the monitoring campaign, we tested whether this finding can be confirmed by constructing a light curve of  photon index, describing the primary X-ray emission. The resulting plot, shown in Fig.~\ref{figure:fig7}, suggests that the time series of the photon index is consistent with the hypothesis of constancy because of the large uncertainties associated with the $\Gamma$ values. This
is indeed confirmed by a $\chi^2$ test, which yields $\chi^2/dof=52.6/80$ ($P_\chi^2=0.99$). Note that the same conclusion is reached using a single BMC model or a power law model to fit the continuum.

\begin{figure}
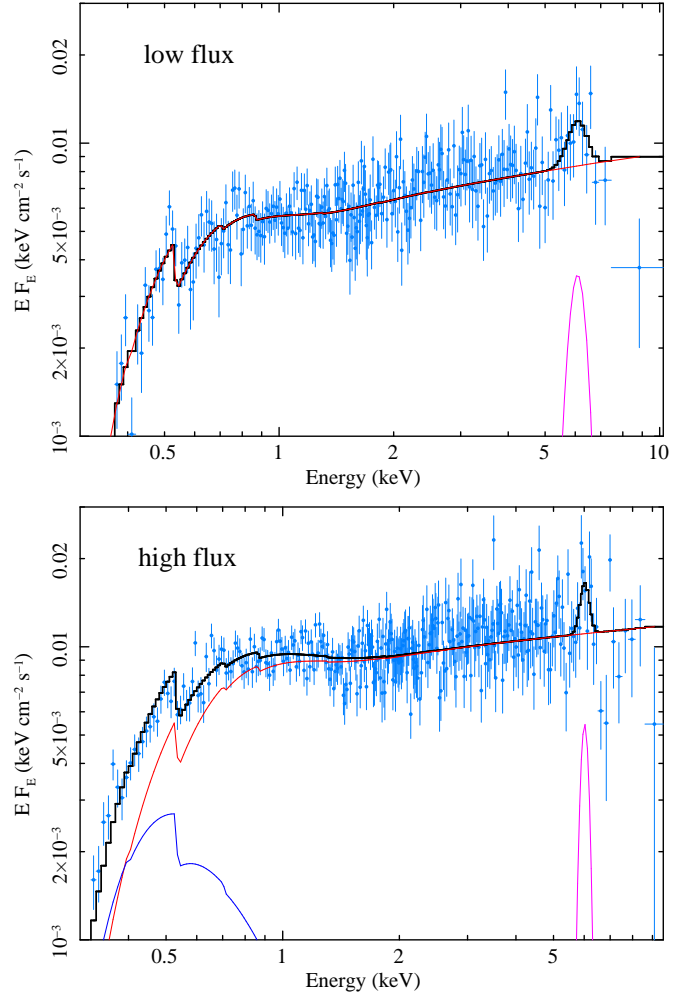

\includegraphics[bb=40 2 600 704,clip=,angle=-90,width=9.cm]{fig8a.eps}
\includegraphics[bb=80 2 600 704,clip=,angle=-90,width=9.cm]{fig8b.eps}
\caption{Unfolded (eeufspec command in Xspec) XRT spectra of \ark\ for the combined low-flux and high-flux observations.}
\label{figure:fig8}
\end{figure}

In an attempt to test whether the spectrum of \ark\ genuinely steepens when the source
brightens, we combined several individual spectra of observations with low count rate
(count rate$_{0.3-10~keV} \leq 0.9$ c/s) to produce a ``low-flux" spectrum, and similarly several spectra with high count rate (count rate$_{0.3-10~keV} \geq 1.4$ c/s) to obtain a ``high-flux" spectrum. The threshold count rate values, shown in the bottom panel of Fig.~\ref{figure:fig1}, were arbitrarily chosen to be distinct from the mean count rate ($1.20\pm0.03$ c/s) and to encompass at least ten individual observations each, in order to increase the S/N of the combined spectra. 
\setcounter{table}{2}
\begin{table} 
\begin{minipage}{180mm}
\caption{Results from the spectral fitting of high- and low-flux of Ark 120}
\begin{tabular}{lcc}
\hline        
\hline
\noalign{\smallskip}
Results & Low-flux & High-flux \\
\noalign{\smallskip}
\hline
\noalign{\smallskip}
$\chi^2$/dof & 1233.5/261 & 404.6/380  \\
$kT_1$ (keV) & $0.13\pm0.01$ & $0.15\pm0.01$\\
$\alpha_1$   & $0.80\pm0.04$ & $0.87\pm0.04$ \\
$\log(A_1)$  &  0.08         & 0.2 \\
$Norm_{BMC1}$     & $(2.6\pm0.1)\times 10^{-4}$ & $(3.5\pm0.1)\times 10^{-4}$\\
$kT_2$ (keV) & $\dots$ & $0.07\pm0.014$\\
$\alpha_2$   & $\dots$ & 4 \\
$\log(A_2)$  & $\dots$ & 0.2 \\
$Norm_{BMC2}$     & \dots & $(3.5\pm0.1)\times 10^{-4}$\\
$E_{\rm line}$ (keV) & $6.3\pm0.1$ & $6.2\pm0.1$\\
$\sigma_{\rm line}$ (keV) & 0.35 & 0.1\\
$Norm_{\rm line}$     & $(8.5\pm2.5)\times 10^{-5}$ & $(5.2\pm2.5)\times 10^{-5}$\\
\noalign{\smallskip}
\hline        
\hline
\end{tabular}
\end{minipage}
\label{tab3}
\vskip 0.1cm 
The errors given in this table are 1-$\sigma$ errors.
\end{table}

The resulting low-flux and a high-flux spectra have well separated mean count rates, $0.77\pm0.02$ c/s and $1.54\pm0.02$ c/s, and net exposures of $\sim$11 ksec and $\sim$12 ksec, respectively.
Importantly, they have considerably higher signal-to-noise ratios compared to individual spectra, as demonstrated by the comparison of Figure~\ref{figure:fig8} with Figure~\ref{figure:fig6}. This allows a better characterization of the spectral models, even though some parameters (such as the Comptonization fraction, or the Gaussian line parameters) remain poorly constrained. 

Restricting the fit to the 2--10 keV range, to avoid complications with the putative soft excess, the high-flux spectrum appears significantly steeper ($\Gamma=1.90_{-0.02}^{+0.04}$) than the low-flux spectrum ($\Gamma=1.72_{-0.06}^{+0.07}$). The difference in slope between the low- and high-flux spectra is seen in Fig.~\ref{figure:fig8}.  Note that for unfolded spectra (i.e., plots of $E F_E$ vs. $E$, which are equivalent to the 
$\nu f_\nu$ vs. $\nu$ plots often used in SED studies of AGN), the slope is given by $2-\Gamma$, which means that the steeper positive slope observed in the low-flux unfolded spectrum (top panel) corresponds to a lower value of $\Gamma$ compared to the high-flux spectrum (bottom panel).

When the 0.3--10 keV range is considered, both low- and high-flux spectra are reasonably well fitted with a coronal BMC model, and both spectra seem to require a Gaussian line ($EW\sim 100-200$ eV), whose addition reduces the $\chi^2$ by 5.4 and 6.5 (for three additional parameters) in the low and high-flux cases, respectively. However, only the high-flux spectrum
requires a second BMC model to fit the soft energy range. The best-fit values of this fitting procedure are summarized in Table~3.

 This is confirmed by the flux-selected spectral fitting analysis that shows that the high-flux spectrum is indeed significantly steeper ($\Gamma=1.90_{-0.02}^{+0.04}$) than the low-flux spectrum ($\Gamma=1.72_{-0.06}^{+0.07}$), and 
that only the high-flux spectrum is statistically improved by the addition of a second BMC model (with $kT\sim0.03$ keV) to fit the low-energy part of the spectrum.
 
In summary, the spectral analysis based on model fitting of flux-selected spectra confirms the existence of spectral variability during the \swift\ campaign of \ark\ revealed by the model-independent spectral variability analysis, with steeper spectra observed when the source has higher count rate.
\section{Correlation analysis}
We used the discrete correlation function (DCF) method of \citet{edel88} to compute the correlation function (CCF) at lags $k = 0,\pm l \Delta t$, where $l = 1,\dots,10$, $\Delta t = 2$ days. As a reference, we used the soft X-ray light curve (hereafter, SX indicates the energy range 0.3-1 keV, and HX the hard X-ray light curve in the 1-10 keV range). We computed the HX vs. SX, M2 vs. SX, U vs. SX, and U vs. M2  cross correlations. In the case of the U vs. M2 correlation, we considered the M2 light curve as reference light curve. Positive lags mean that the reference light curve leads, negative lags indicate that the reference light curve follows. We calculate the centroid of the DCF, $\tau_{\rm cent}$, as the mean of all the DCF points which are $>$ 0.75$\times$DCF$_{\rm max}$, and we accept it as our estimate of the time-lag between two light curves. We also compute the average DCF$_{\rm max}$ as the mean of the DCF values of the same points. 

\begin{figure*}
\begin{center}
\includegraphics[bb=0 35 545 760,clip=,angle=0,width=13.cm]{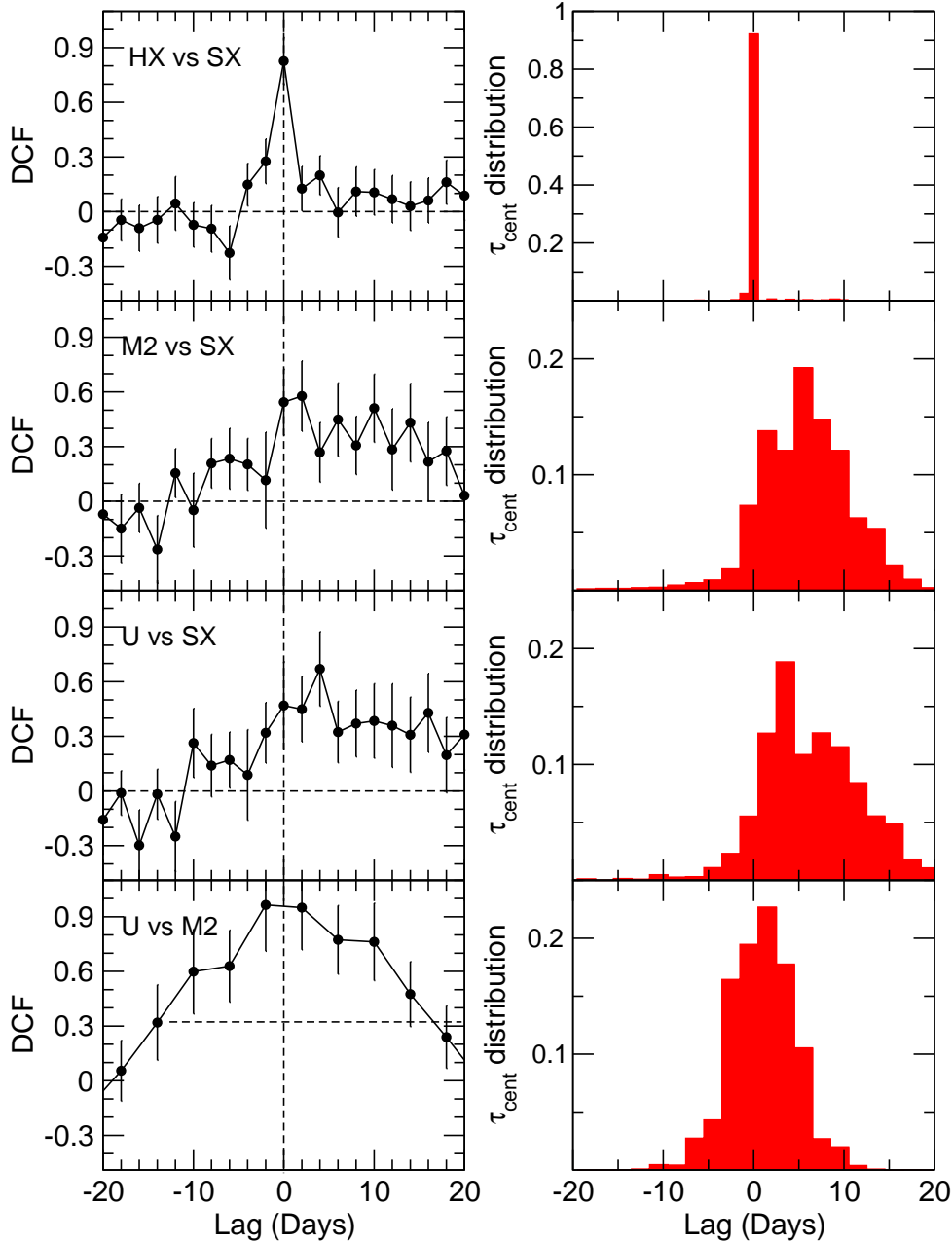}
\caption{{\it Left panels:} plots of cross-correlation between the 0.3--1 keV soft X-ray flux and the hard (1--10 keV) X-rays (top panel), UVM2 flux (second panel), U flux (third panel), and U vs. UVM2 (bottom panel). {\it Right panels:} distributions of the centroid time lags for the various cross correlations.} 
\label{figure:fig9}
\end{center}
\end{figure*}

The resulting CCFs are shown in the left panels of Fig.~\ref{figure:fig9}. The HX vs. SX shows a strong, narrow peak at zero lag. On the other hand, the UV/X-ray correlations are skewed towards positive lags, suggesting that the SX band variations lead those in the UV band. DCF$_{\rm max}$ values are smaller in the cross-correlations between the UV light curves and the X-ray band. This is not surprising, given the fact that the UV band light curves are much ``smoother" than the X-ray band light curves (see Fig.~\ref{figure:fig1}). Finally, when U is cross-correlated with the  UVM2 band, the CCF is roughly symmetric and shows a strong peak of the order of DCF$_{\rm max}\sim 0.9$. 

In Table~\ref{tab4}, we list DCF$_{\rm max}$ together with the the time-lags, $\tau_{\rm cent}$, between the various bands along with their respective 90\% errors. The errors were estimated using the Monte Carlo simulation method proposed by \citet{peter98}. For each light curve pair that we cross-correlated, we produced 10000 simulated light curves following their “random subset selection” prescription.  We computed the DCF of each light curve pair, $\tau_{\rm cent}$, and DCF$_{\rm max}$ exactly as we did with the observed light curves. We used the 10000 values to build up the $\tau_{\rm cent}$, and DCF$_{\rm max}$ distribution function. The distribution of the centroid time lags are also plotted in Fig.~\ref{figure:fig9} (right panels). We used these distributions to estimate the 90\% confidence limits, which we assume are representative of the 90\% confidence limits of the computed $\tau_{\rm cent}$ and DCF$_{\rm max}$ values when using the observed light curves. 

We do not find a statistically significant detection of delays between any of the light curve pairs considered: all the “lags” listed in Table~\ref{tab4} are consistent with zero within their 90\% confidence limits. In the case of the Hard vs. Soft X--ray, the delays are very small  (variations in HX and SX happen almost simultaneously). On the other hand, positive values of $\tau$ of the order of a few ($\sim 4$) days are tentatively detected in the case of the cross-correlation between the UV light curves and the soft X-ray band. The lags are identical when we consider the cross-correlation between the UVM2 and the U band light curves. 

\begin{figure*}
\begin{center}
\includegraphics[bb=30 460 520 760,clip=,angle=0,width=12.cm]{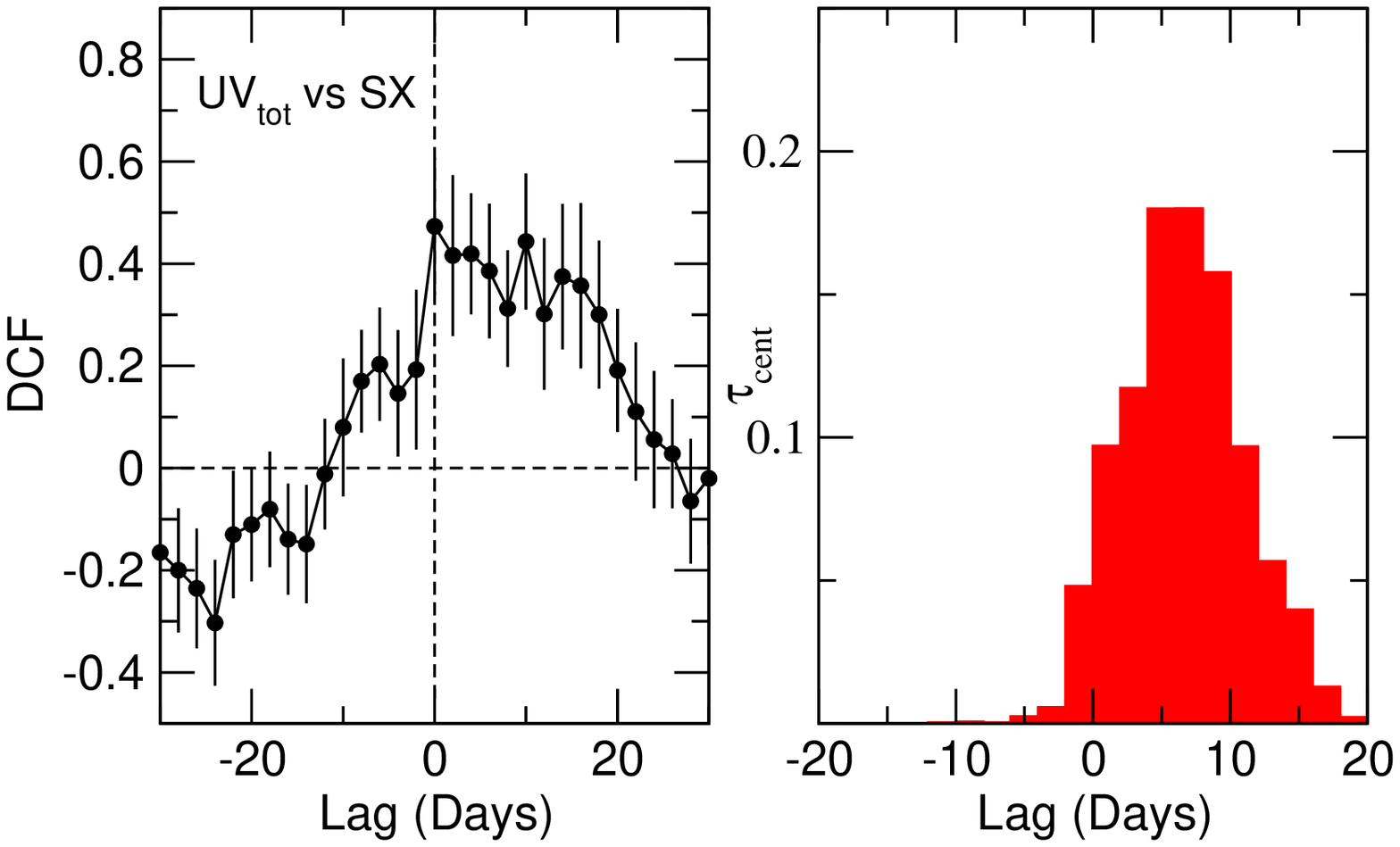}
\caption{{\it Left panel:} Plots of cross-correlation between the combined (U+UVM2) light curve and the soft X-ray energy band 0.3--1 keV. {\it Right panel:} distribution of the centroid time lags for this cross correlations.} 
\label{figure:fig10}
\end{center}
\end{figure*}

In order to reduce the statistical uncertainty in this cross-correlation analysis and better constrain the delay between X-ray and UV light curves in \ark, we
tried to combine the U and UVM2 light curves, which have similar trends and do not show any significant delay. To this end, we first interpolated the U light curve, and then shifted the UVM2 light curve by a multiplicative factor obtained by minimizing the RMS between the interpolated U values and the shifted UVM2 values. The resulting combined light
curve (hereafter, U+UVM2) is simultaneous to the X-ray light curve and has the same number of data points. The CCF analysis with this combined light curve, shown in Fig.~\ref{figure:fig10}, reveals that the soft X-ray light curve leads the UV one by $7.5\pm7$ days (errors indicate the 90\% confidence limits). 
The results are not affected by the uncertainty in the scaling factor when we create the combined  U+UVM2 light curve, due to the small uncertainty in the $F_{\rm var}$ of the individual light curves.

For completeness, we also performed a cross-correlation between the SX and the X-ray hardness ratio, $HR$, and between the combined UV light curve and $HR$. In both case, the flux light curves were the reference light curves. The results of this analysis 
indicate that $HR$ is anti-correlated with both SX and (U+UVM2).

In summary, our cross correlation analysis confirms that soft and hard X-ray variations are strongly correlated and occur nearly simultaneously. Similarly, the variations in the U and UVM2 filters appears to be correlated and without substantial delay. When the UV light curves are correlated with the X-ray light curve, a possible (but not statistically significant) delay is suggested, with the X-ray leading the variations in the UV bands by a few days. This result is confirmed at a higher significance level when the U and UVM2 are combined and then correlated with the soft X-ray. 

\setcounter{table}{3}
\begin{table} 
\footnotesize
\caption{UVOT Correlation analysis results}
\begin{center}
\begin{tabular}{ccc} 
\hline        
\hline
\noalign{\smallskip}
Energy bands    & $\tau_{\rm cent}$ (d) & DCF$_{\rm max}$\\
(1) & (2) & (3)\\
\noalign{\smallskip}
\hline 
\noalign{\smallskip}
HX vs. SX  &   $0\pm 0.5$ & $0.83^{+0.17}_{-0.13}$\\
\noalign{\smallskip}
\hline
\noalign{\smallskip}
UVM2 vs. SX &    $4.5^{+9.5}_{-6.5}$ & $0.52^{+0.45}_{-0.02}$\\
\noalign{\smallskip}
\hline
\noalign{\smallskip}
U vs. SX  &   $4^{+12}_{-5.5}$ & $0.67^{+0.30}_{-0.08}$ \\ 
\noalign{\smallskip}
\hline
\noalign{\smallskip}
U vs. UVM2  &   $4^{+3}_{-8}$ & $0.86^{+0.15}_{-0.07}$ \\
\noalign{\smallskip}
\hline
\noalign{\smallskip}
(U+UVM2) vs. SX  &  $7.5\pm 7$ & $0.52^{+0.24}_{-0.12}$\\
\noalign{\smallskip}
\hline
\hline
\noalign{\smallskip}
HR vs. SX  &   $0^{+1}_{-10}$ & $-0.79^{+0.26}_{-0.21}$ \\
\noalign{\smallskip}
\hline
\noalign{\smallskip}
HR vs. (U+UVM2)  &   $-15^{+8}_{-6}$ & $-0.48^{+0.39}_{-0.08}$ \\
\noalign{\smallskip}
\hline        
\hline
\end{tabular}
\end{center}
{\bf Columns Table 4}: 1= Correlated light curves. 
2= Lags measured in days with the 90\% errors. 3= Maximum of DCF with the 90\% errors.
\label{tab4}
\footnotesize
\end{table}

\section{Summary and Conclusion} 
We first summarize the most relevant results of our \swift\ monitoring campaign of 
\ark, and then discuss their implications in the broader context of AGN variability studies.

\begin{itemize}
\item {\it Temporal variability:}
The 6-month XRT and UVOT monitoring of \ark\ revealed that strong variability in the X-ray and UV bands, observed in past pointing observations on shorter timescales, occurs on all timescales probed by the \swift\ campaign, i.e., from a few days to a few months (see Fig. ~\ref{figure:fig1} and Fig.~\ref{figure:fig2}). While the X-ray variability is characterized by frequent large-amplitude changes where the count rate can double or halve in periods as short as 2--4 days, the variations observed in the UV bands are smoother with flux changes of the order of 20-40\% occurring on timescales of months.
This different behavior can be quantified by fractional variability measurements: $F_{\rm var}$ increases from $\sim 5\%$ in the U band, to $\sim 8\%$ in UVM2, up to $\sim 20\%$
in the 0.3--10 keV energy band.

\item {\it Spectral variability:}
The continuous temporal variability of \ark\ appears to be associated with persistent spectral variability based on various model-independent analyses. For example, the light curve of the hardness ratio $HR=(h-s)/(h+s)$ is inconsistent with the hypothesis of constancy at a high confidence level. When $HR$ is plotted vs. the total X-ray count rate, a weak but
statistically significant anti-correlation is found, indicating that the X-ray spectrum softens when the source brightens, which is the typical behavior observed in Seyfert galaxies (see Fig.~\ref{figure:fig4}). Additionally, soft and hard X-ray fluxes are tightly correlated and well described by a linear equation, whose slope is consistent with unity and whose intercept is inconsistent with zero, suggesting the presence of a constant hard component. Finally, the light curve of the broadband spectral index indicates that the entire SED varies throughout the monitoring campaign (see Fig.~\ref{figure:fig5}).

\item {\it Spectral analysis:} 
The spectral analysis of individual $\leq$ 1 ks observations does not provide conclusive results about the long-term spectral variability of \ark, due to the limited statistics.
However, combining several individual spectra into a low-flux and a high-flux spectrum and
then performing a model fitting of these two flux-selected spectra makes it possible to conclude that the steeper-when-brighter behavior is caused by the steepening of the photon index.

\item {\it Correlation analysis:}
A cross correlation analysis of the \swift\ UVOT and XRT light curves of \ark\ indicates 
that soft and hard X-ray variations are strongly correlated and occur nearly simultaneously. Also the U and UVM2 light curves are well correlated with each other and do not show any substantial delay. When the UV light curves are correlated with the X-ray light curve, a possible (but not statistically significant) delay is tentatively detected, with the X-ray leading the variations in the UV bands by a few days (see Fig.~\ref{figure:fig9}). This result is confirmed at a higher significance level by using the combined U and UVM2 light curve for the correlation analysis with the soft X-ray light curve (see Fig.~\ref{figure:fig10}). 
Finally, the hardness ratio $HR$ appears to be anti-correlated with the soft X-ray light curve, and with the combined UV light curve. While there is no relevant lag between $HR$ and X-ray flux, it appears that the changes in the UV light curve are delayed by several days ($\tau=-15_{-6}^{+8}$ d) with respect to the $HR$ changes. 
The reason for the UV - HR correlation was to investigate the possibility that the observed UV photons are the input soft photons up-scattered in the hot corona. In this case, an increase in the flux of soft photons may cause the cooling of the corona, and hence a steepening in the observed X-ray spectrum, as it has been observed in the past (e.g., \citealt{nan00}). Our results do not support this possibility and may be explained by the fact that the X-rays and HR are strongly anti-correlated (with no delay), and the X-rays and UV are moderately correlated with a delay of a few days.

\end{itemize}

Our study confirms that \ark\ behaves as a typical Seyfert galaxy with persistent X-ray
(and UV) flux variability associated with spectral variability, where the spectrum softens as 
the source brightens. This spectral behavior too is common to the vast majority of Seyfert
galaxies. However, the ``bare" nature of \ark\ ensures that the spectral variability is
caused by intrinsic changes in the primary emission, rather than being associated with variations of the absorber surrounding the source, as suggested for many other AGN.
Our results, obtained from a model-independent analysis of monitoring data spanning several months, appear to be consistent with those based on detailed spectral analysis of broadband spectra obtained from the long uninterrupted exposures. For instance, the fact that the
intercept of the hard vs. soft X-ray correlation is positive and inconsistent with zero
indicates the presence of a constant hard component, which is naturally explained by the
reflection component detected by \citet[]{matt14} using high-quality spectra from \xmm\ and {\it NuSTAR}.

The results from the cross-correlation analysis (the tentative time lag of the UV flux 
with respect
 to the X-ray light curve) are consistent with the reprocessing scenario, where changes in
the UV/optical emitting accretion disk are driven by changes in the X-ray corona. Although the measured time lag is poorly constrained due to the large statistical uncertainty ($\tau=7.5\pm7$ d, which is obtained when the
combined UV light curve is used for the correlation analysis), to put \ark\ in context, it is helpful to compare its correlation results with those obtained in similar studies.   

\begin{figure}
\begin{center}
\includegraphics[bb=30 30 350 300,clip=,angle=0,width=8.5cm]{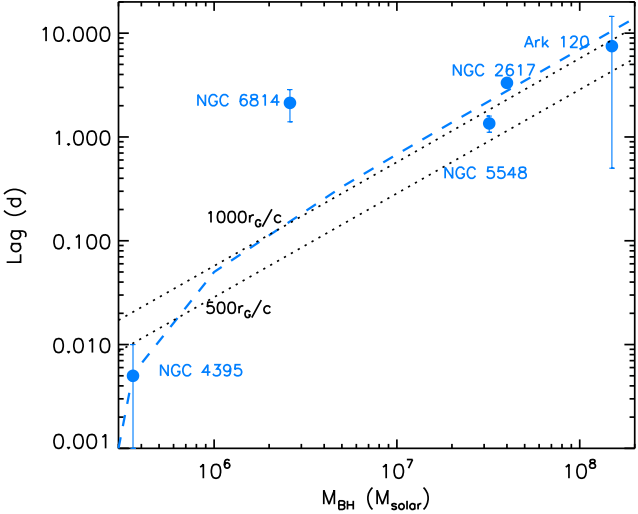}
\caption{Time lags of the UV band with respect to the X-ray band plotted vs. the black hole mass. The dashed line represents the best-fit linear model $UV/X-ray~Lag=-0.02\pm0.01+(7.1\pm0.4)\times10^{-8}M_{\rm BH}$. The dotted dark lines represent the light travel times for 500$r_G/c$ (bottom) and 1000$r_G/c$ (top).} 
\label{figure:fig11}
\end{center}
\end{figure}
In particular, it is interesting to investigate whether there
exists a correlation between time lags and $M_{\rm BH}$ or $\dot m$, using AGN whose UV and X-ray light curves have been simultaneously monitored by \swift\ for several months. In addition to the two objects studied by our group:
PKS 0558-504 --$M_{\rm BH}\sim3\times10^8 ~{\rm M_\odot}$ and $\dot m\geq 1$-- \citep{glioz13}, and \ark\ --$M_{\rm BH}=1.5\times10^8 ~{\rm M_\odot}$ and $\dot m=0.005$-- (this work), this limited sample of AGN comprises NGC 4395 --$M_{\rm BH}=3.6\times10^5 ~{\rm M_\odot}$ and $\dot m=0.005$-- \citep{came12}, NGC 2617 --$M_{\rm BH}=4\times10^7 ~{\rm M_\odot}$ and $\dot m\sim0.1$-- \citep{shap14}, NGC 5548 --$M_{\rm BH}=3.2\times10^7 ~{\rm M_\odot}$ and $\dot m=0.03$-- \citep{edel15}, and  NGC 6814 --$M_{\rm BH}=2.6\times10^6 ~{\rm M_\odot}$ and $\dot m=0.01$-- \citep{panc14,troy16}.

In Figure~\ref{figure:fig11} we plotted the time delays between the UV band and the X-ray detected for these objects. For all objects, we used the lag value reported for the UVOT U filter ($\lambda_{\rm peak}=350$ nm), with the exception of NGC 6814 for which only the lag of UVW1 ($\lambda_{\rm peak}=260$ nm) was measured.  PKS 0558-504 was not included because the detection of a lag was not statistically significant. However, it is worth noting that PKS 0558-504 putative delay
($\tau=-16.8_{-14.7}^{+16.8}$ days) was of the same order as the one detected in \ark, which has a similar black hole mass, but negative (i.e., with the UV emission that appears to lead the X-rays by a few days). It is
interesting to note that PKS 0558-504 is the only object for which such a negative delay has
been putatively detected and the only AGN of this sample with accretion rate above the Eddington level.

All objects in Fig.~\ref{figure:fig11}, with the exception of NGC 6814, are reasonably well fitted with a linear model, represented by the dashed line $UV/X-ray~Lag=-0.02\pm0.01+(7.1\pm0.4)\times10^{-8}M_{\rm BH}$. This finding is qualitatively in
agreement with the general picture of black hole systems, where the length-scale is set by
the black hole mass, naturally implying a larger physical separation (and hence longer delays) for systems with larger $M_{\rm BH}$.

The dotted lines, which represent
the light travel times for 500 $r_G/c$ and 1000 $r_G/c$, suggest that, with the exception of NGC 4395 (for which no significant lag was detected), all AGN of this sample require a physical separation between the X-ray emitting region and the UV region of the order of 1000 $r_G$ or more (NGC 6814). These values are considerably larger than the physical locations of the UV emitting region predicted by the standard accretion disk model; using Equation (2) from \citealt{came12}) we obtain values of the order $100-250 r_G$. We therefore conclude that, for this sample of AGN, these cross-correlation results imply a larger accretion disk compared to the Shakura-Sunyaev standard model, as suggested by recent findings based on micro-lensing studies from \citet{mosq13} and intensive simultaneous monitoring of several energy bands in NGC 5548 \citep{edel15}.

In conclusion, our work indicates that long-term monitoring studies of AGN 
provide useful information which is complementary to that obtained in long-exposure spectral studies. Importantly, combining \ark\ correlation results with those of similar
studies of AGN monitored by \swift, suggests the existence of a positive correlation between time lags and 
black hole mass. Additional monitoring studies of AGN spanning a broader range of $M_{\rm BH}$ and ${\rm M_\odot}$ are necessary to derive a firmer conclusion.\\

\noindent{\bf \Large Acknowledgements}\\ 
We thank the anonymous referee for the constructive commentss that have improved the clarity of the paper. MG acknowledges support by the SWIFT Guest Investigator Program under NASA grant NNX15AB64G. 

\end{document}